# Rectangular SNAP microresonator fabricated with a femtosecond laser

Qi Yu,[1,2,*] Sajid Zaki,[1] Yong Yang,[1] Nikita Toropov,[1] Xuewen Shu,[2] and Misha Sumetsky[1]

[1]Aston Institute of Photonic Technologies, Aston University, Birmingham B4 7ET, UK
[2] Wuhan National Laboratory for Optoelectronics & School of Optical and Electronic Information, Huazhong University of Science and Technology, Wuhan 430074, China
*Corresponding author: yqhtyz@163.com

**SNAP microresonators, which are fabricated by nanoscale effective radius variation (ERV) of the optical fiber with sub-angstrom precision, can be potentially used as miniature classical and quantum signal processors, frequency comb generators, as well as ultraprecise microfluidic and environmental optical sensors. Many of these applications require the introduction of nanoscale ERV with a large contrast α which is defined as the maximum shift of the fiber cutoff wavelength introduced per unit length of the fiber axis. The previously developed fabrication methods of SNAP structures, which used focused $CO_2$ and femtosecond laser beams, achieved α ~ 0.02 nm/μm. Here we develop a new fabrication method of SNAP microresonators with a femtosecond laser which allows us to demonstrate a 50-fold improvement of previous results and achieve α ~ 1 nm/μm. Furthermore, our fabrication method enables the introduction of ERV which is several times larger than the maximum ERV demonstrated previously. As an example, we fabricate a rectangular SNAP resonator and investigate its group delay characteristics. Our experimental results are in good agreement with theoretical simulations. Overall, the developed approach allows us to reduce the axial scale of SNAP structures by an order of magnitude.**

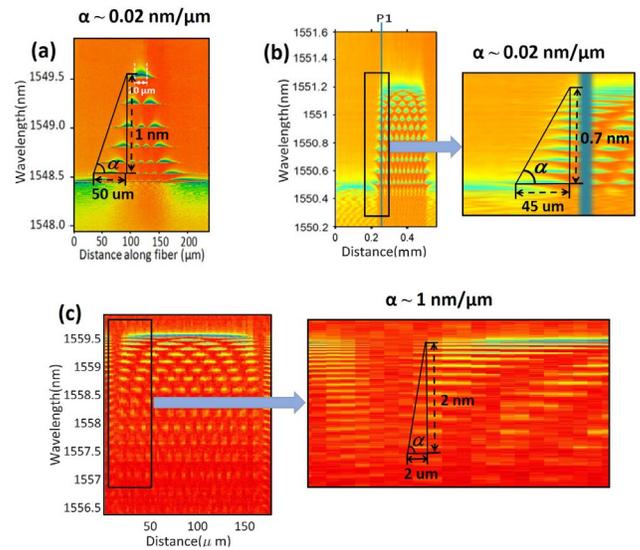

**Fig. 1.** Estimation of the contrast $\alpha$ achieved in SNAP structures introduced by (a) annealing with a $CO_2$ laser, (b) axially oriented lines inscribed with a femtosecond laser, and (c) the approach of the present Letter.

SNAP (Surface Nanoscale Axial Photonics) is an emerging nanophotonic technological platform which enables the fabrication of miniature WGM resonant photonic circuits at the surface of an optical fiber with unprecedented sub-angstrom precision [1, 2]. A SNAP microresonator supports whispering gallery modes (WGMs) which circulate near the fiber surface and slowly propagate along its axis. Multifunctional SNAP devices can be created by modifying a uniform fiber with the introduction of nanoscale effective radius variation (ERV). SNAP resonant structures with nanoscale ERV can be created by using different approaches including annealing with focused $CO_2$ laser beams [1, 2], femtosecond laser inscription [3, 4], local heating [5], tapering [6], bending [7], and evanescent coupling with droplets in microcapillaries [8].

One of the important applications of the SNAP platform consists in fabrication of miniature processors of optical pulses. For example, a ground breaking miniature SNAP delay line with semi-parabolic nanoscale ERV $\Delta r(z) \sim -z^2$ along the fiber axis $z$ demonstrated the dispersionless group delay time (~2.58 ns) of 100 ps pulses with low insertion losses (< 1.12 dB/ns) [2]. This resonator can be potentially employed as an optical buffer, which can trap and release optical pulses without distortion [9]. As another example, the SNAP resonator with the ERV $\Delta r(z) \sim -z^{2/3}$ experimentally demonstrated in Ref. [10] presented a miniature device enabling 20 ns/nm dispersion compensation of 100 ps pulses. It had been suggested in Ref. [2] that, to arrive at the maximum operational bandwidth and best impedance matching between a SNAP microresonator delay line and other signal processors, it is necessary to maximize the steepness of ERV at the microresonator edge where the input-output microfiber is positioned.

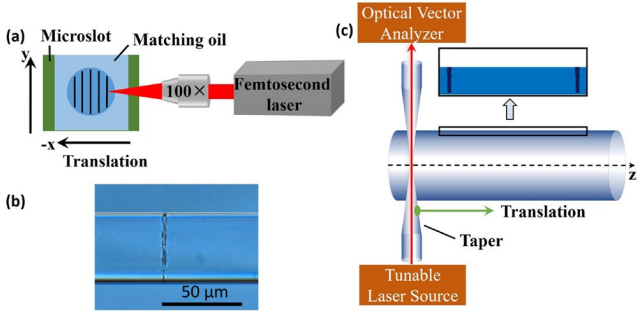

**Fig. 2.** (a) Illustration of the setup for the fs laser fabrication. (b) Optical microscope image of an ICS. (c) Characterization of the introduced rectangular SNAP resonator; inset: illustration of two ICSs.

Here we note that the problem of increasing the steepness (or contrast) of ERV is of general importance because its solution allows us to fabricate SNAP structures having much smaller dimensions. Since the spectrum of the structure rather than its ERV is of our major concern, we determine its contrast $\alpha$ as the maximum shift of the WGM cutoff wavelength $\Delta\lambda_c(z)$ which is introduced per unit fiber length and express $\alpha$ in nm/μm. The ERV $\Delta r(z)$ is related to the shift $\Delta\lambda_c(z)$ as $\Delta r(z) = \Delta\lambda_c(z) r_0 / \lambda_c$, where $r_0$ is the fiber radius. The contrast $\alpha$ can be simply determined from the spectrogram of the SNAP structure, which is the surface plot of its transmission spectra $T(z,\lambda)$ measured as a function of wavelength $\lambda$ and coordinate $z$ along the SNAP structure axis using an input-output microfiber (see Fig. 2(c) for the illustration of our measurement setup) [11, 12]. As an example, Fig. 1(a) shows the spectrogram of a SNAP microresonator introduced by annealing with a focused $CO_2$ laser beam in Ref. [1]. The contrast of this resonator is $\alpha \sim 0.02$ nm/μm. As another example, Fig. 1(b) shows the spectrogram of a SNAP microresonator introduced by axially oriented lines inscribed with a femtosecond laser in Ref. [3]. The contrast of the introduced microresonator determined from the magnified inset of this figure is also $\alpha \sim 0.02$ nm/μm. To the best of our knowledge, Figs. 1(a) and 1(b) demonstrate the largest characteristic contrasts achieved to date in SNAP technology. For $CO_2$ laser annealing, the maximum demonstrated cutoff wavelength shift is $\sim 1$ nm and the contrast is limited by the characteristic size of the laser beam $\sim 50$ μm. For femtosecond laser inscription in the fiber core [3, 4], the size of the laser beam can be as small as a micron and the maximum cutoff wavelength shift is, similarly, close to 1 nm. However, the laser-induced stresses inside the fiber spread the effect of the laser inscription along the fiber length having the order of the fiber radius.

The idea of the present Letter is to overcome the contrast limitations of the methods developed previously with an alternative approach employing a femtosecond laser. Our new method described below allowed us to achieve the contrast of $\alpha \sim 1$ nm/μm, i.e., demonstrate a 50-fold improvement of the results achieved previously (see the rectangular microresonator in Fig. 1(c) and its inset).

In our experiments, an amplified Ti: sapphire laser (800 nm wavelength) with $\sim 110$ fs pulse duration and 1 kHz repetition rate was used to inscribe lines at the cross-section of optical fibers with cladding radius $r_0 = 19$ μm. Each fiber was stripped from its protective coating and immersed in a microslot filled with refractive index matching oil to compensate for the lensing effect of the curved fiber surface. The laser beam was focused by a 100× objective (NA = 0.55). A high-resolution translation stage was used to translate the fiber relative to the laser beam with a speed of $\sim 5$ μm/s. As shown in Fig. 2(a), firstly, the focused laser beam was located on the left surface of the fiber. Then the fiber was translated along the +y axis to achieve each inscribed line. Afterwards, the fiber was translated along the -x axis in order to change the focused location of the laser beam, and then a new line was inscribed along +y axis. The above process was iterated until the focus position of the laser beam reached the right surface of the optical fiber. In order to ensure complete coverage of the fiber cross-section, the translation distance along the y-axis was made slightly larger than the fiber diameter. Specifically, each inscribed cross-section (ICS) employed in our experiments was introduced by 12 lines separated with 3 μm. The optical microscope image of an ICS having a micron-scale axial width is shown in Fig. 2(b).

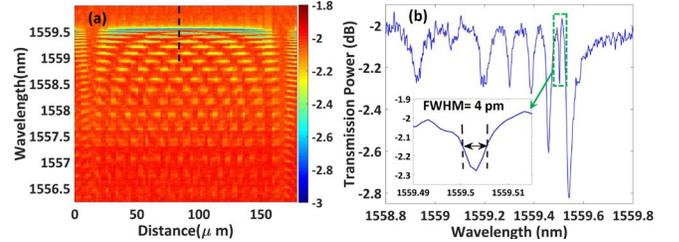

**Fig. 3.** (a) Spectrogram of the rectangular SNAP resonator with 140 μm along the fiber axis. (b) Transmission spectrum of this resonator at the position z =78 μm marked by a black line in (a). The inset of (b) shows the FWHM of the second order axial mode marked with a green dashed rectangle.

As shown below, the two ICSs schematically illustrated in Fig. 2(c) form a rectangular SNAP resonator between them. The axial length of this resonator can be controlled by the distance between these ICSs.

To estimate the effect of losses introduced by ICSs a rectangular resonator with axial length 140 μm was fabricated. The spectrogram $T(z,\lambda)$ of this resonator shown in Fig. 3(a) was measured using a biconical taper with a micron-diameter waist (microfiber) positioned transversely to the resonator and connected to the LUNA Optical Vector Analyzer (OVA) with 1.3 pm spectral resolution (Fig. 2(c)). The microfiber was translated along the resonator and touched it periodically with 2 μm steps. It is seen from Fig. 3(a) that the bandwidth, where the resonator spectrum is resolved, is close to 3 nm corresponding to the ERV of $\Delta r = \Delta\lambda \cdot r_0 / \lambda = 36$ nm. The contrast achieved at the edge of this resonator is estimated from the magnified spectrogram shown in

Fig. 1(c). From this spectrogram, the change of ∼ 2 nm in the value of cutoff wavelength is introduced along the axial length of around 2 µm. Therefore, we estimate the achieved contrast as 1 nm/µm. A better spatial resolution measurement can give us a better estimate of the introduced contrast, which can exceed 1nm/µm. It is seen from Fig. 3(a) that the Q-factor of WGMs degrades with the spectral separation from the cutoff wavelength (the top of the resonator). This can be explained by increasing of the tunneling leakage and scattering of light through the ICS as this separation grows. The second axial mode of this resonator with wavelength ∼ 1563.68 nm was chosen to estimate the intrinsic Q factor of the resonator. The transmission spectrum of this axial mode at contact position $z = 78$ µm is shown in Fig. 3(b). The inset of Fig. 3(b) shows that the FWHM of the selected resonant mode was ∼ 4 pm corresponding to the intrinsic Q factor of ∼ $4 \times 10^5$.

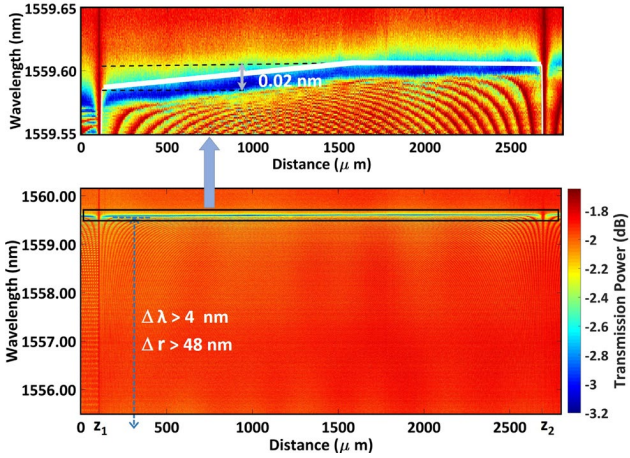

**Fig. 4.** Spectrogram of the rectangular SNAP resonator with 2.58 mm axial length. Inset: a section of this spectrogram magnified near the cutoff wavelength.

As another verification and an application of the developed method, we have fabricated a longer rectangular microresonator and investigated its group delay time characteristics. Theoretically, group delay in a rectangular SNAP resonator was studied in the Supplemental Material to Ref. [2]. However, to the best of our knowledge, this problem has not been addressed experimentally. For this purpose, we introduced a rectangular SNAP resonator by setting the distance between two ICSs equal to 2.58 mm and keeping other fabrication parameters the same as those of the resonator shown in Fig. 3(a). In order to reduce the scattering losses, the resonator was polished by a $CO_2$ laser beam moving with the speed of 0.5 mm/s along the fiber axis. Figure 4 shows the transmission spectrogram of the fabricated resonator. It is seen that the spectral depth of this resonator exceeds 4 nm corresponding to the ERV of greater than 48 nm. While larger ERV was demonstrated for SNAP resonators with smooth ERV [6], to the best of our knowledge, this is the largest high contrast ERV demonstrated to date.

In our experiment, light was coupled into the resonator from the input microfiber at contact point $z_c$ positioned close to the left hand side of the resonator. The input WGMs slowly propagate along the fiber axis and return to the microfiber after multiple reflections from the resonator edges. The interference between reflected waves causes oscillations in the spectrum of the output light. Remarkably, the impedance matching condition can be achieved by adjusting the coupling conditions of this resonator with the input-output microfiber taper so that these oscillations are suppressed, i.e., light fully returns back into the microfiber after a single reflection from the turning points [2].

In order to optimize the coupling conditions and arrive at the impedance matching condition, the taper was translated along its axis and along the axis of the resonator in the vicinity of edge $z_1$. The results of optimization are summarized in Fig. 5(a)-(d). The spectrograms shown in Fig. 5(a) and (b) were obtained by the measurement of Jones matrix spectra with 2 µm steps along the resonator followed by numerical separation of the polarization states by diagonalization of Jones matrices using the approach described in [13]. The position of the microfiber corresponding to the impedance matching condition indicated by the black dashed lines in Figs. 5(a) and 5(b) was found at $z$ = 38 µm, which is about 16 µm away from the resonator edge. At this position, the impedance matching condition was realized at the resonance wavelength ∼ 1563.2 nm. The transmission spectrum and group delay spectrum at this position are shown in Figs. 5(c) and 5(d), respectively. It is seen that the transmission power and group delay ripples are minimized at wavelength ∼1563.2 nm and are relatively small in the neighborhood of this wavelength. The average group delay grows from 0.8 ns to 2.86 ns within the bandwidth from ∼ 1563.0 nm to ∼ 1563.25 nm indicating the strong dispersion of the rectangular SNAP resonator delay line. The insertion loss of this delay line ∼ 6 dB found from Fig. 5(c) is primarily caused by scattering losses in the region of coupling with the microfiber and intrinsic loss of the device.

We compared the experimentally determined delay time of our resonator with the classical theory. In the classical approximation, the group delay of light, which is launched into a SNAP resonator at wavelength $\lambda$ from the input-output microfiber positioned at its edge $z_1$ and returns back after a single round trip, is [2]:

$$\tau(\lambda) = \frac{\lambda_c^2}{\pi c} \int_{z1}^{z2} \frac{\partial \beta(\lambda, z)}{\partial \lambda} dz \quad (1)$$

where the propagation constant

$$\beta(\lambda, z) = \frac{2^{3/2} \pi n_f}{\lambda_c^{3/2}} \left( \lambda_c(z) - \lambda + i\gamma_c \right)^{1/2}. \quad (2)$$

is expressed through the refractive index of the optical fiber $n_f$ and intrinsic resonance width $\gamma_c$, which determines the attenuation of the light in the optical fiber. This value was estimated from the inset in Fig. 3 as $\gamma_c \sim$ 4 pm. For an ideal rectangular resonator, $\lambda_c(z)$ does not depend on $z$ and Eqs. (1), (2) yield $\tau(\lambda) = 2^{1/2} c^{-1} n_f \lambda_c^{1/2} (\lambda_c - \lambda)^{-1/2} (z_2 - z_1)$. However, for small propagation constant of our concern, the WGM wavelength $\lambda$ is very close to the cutoff wavelength $\lambda_c$. Then, even very small variation $\Delta\lambda_c(z)$ can strongly affect the delay time. In order to take this variation into account, we approximated $\Delta\lambda_c(z)$ by the partly linear and partly constant dependence shown by the white line in the inset of Fig. 4, which magnifies the spectrogram of our resonator in the vicinity of $\lambda_c(z)$. It is seen from Fig. 5(d) that the experimental average group delay (red line) is in excellent agreement with the theoretical group delay (black dashed line) found from Eqs. (1) and (2).

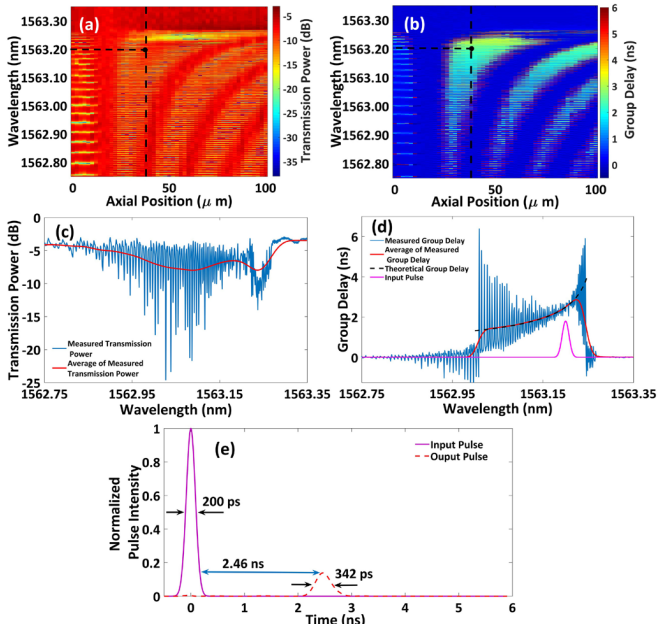

**Fig. 5.** Spectrograms of (a) the transmission power and (b) group delay of the rectangular resonator. The black dashed lines in (a) and (b) shows the impedance matching position. (c) Transmission power spectrum and (d) group delay spectrum at the contact position $z = 38$ μm. The red lines in (c) and (d) shows the average values of transmission power and group delay, respectively. The dashed black curve in (d) depicts the theoretical average group delay of this resonator. (e) The output signal intensity (red dashed line) calculated for the input 200 ps pulse (magenta dashed line) based on the experimental spectra in (c) and (d). The input pulse spectrum is determined by the magenta solid line in (d).

Finally, Fig. 5(e) shows the time-dependent transformation of an optical pulse calculated based on the experimental transmission power and group delay time depicted in Fig. 5(c) and (d). Specifically, we considered a Gaussian pulse with FWHM 200 ps and the central wavelength of ~ 1563.2 nm. The input and output pulses are shown by the magenta solid line and red dashed line in Fig. 5(e), respectively. It is seen that the output pulse is delayed by 2.46 ns, which is close to the value of 2.5 ns of the average group delay at wavelength ~ 1563.2 nm found from the experimental group delay spectrum in Fig. 5(d). Due to the strong dispersion of the rectangular resonator, the FWHM of the output pulse is broadened from 200 ps to ~ 340 ps. The decreased peak intensity of the output pulse is caused by the transmission losses of the device and pulse broadening.

In conclusion, we have developed a new technique for the fabrication of rectangular SNAP microresonators, which consists in the inscription of series of lines along the cross sections of an optical fiber with a femtosecond laser. The steepness of edges of the fabricated resonators is ~ 1 nm/μm which is a 50-fold improvement compared to methods developed previously. As an application, we fabricated a miniature delay line based on the rectangular SNAP resonator and investigated its delay time and dispersion. Further development of the approach demonstrated in this Letter will include the optimization of the femtosecond laser inscription (in particular, the density and geometry of lines and the power of laser) as well as combination of this approach with other fabrication methods of SNAP microresonators. Overall, the developed approach allows us to reduce the axial scale of SNAP structures by an order of magnitude.

**Funding.** Engineering and Physical Sciences Research Council (EPSRC) (EP/P006183/1); Horizon 2020 MCSA COFUND MULTIPLY (H2020 GA 713694); National Natural Science Foundation of China (NSFC) (61775074); National Key R&D Program of China (SQ2018YFE010506).

**Acknowledgment.** Q. Y. acknowledges the China Scholarship Council for financial support. Q. Y. is grateful to Kaiming Zhou and Neil Gordon for consultations regarding the operation of the femtosecond laser used in this work.

**Disclosures.** The authors declare no conflicts of interest.